\documentclass{emulateapj}

\newcommand{\mjybm}{\mbox{mJy~beam${}^{-1}$}}

\shortauthors{LaRosa et al.}
\shorttitle{6~cm study of NTFs}
\begin{document}
Astrophysical Journal, Accepted

\title{New Nonthermal Filaments at the Galactic Center: Are They Tracing a Globally Ordered Magnetic Field?}

\author{T.~N.~LaRosa}
\affil{Department of Biological \& Physical Sciences,
 Kennesaw State University, 1000 Chastain Rd., Kennesaw, GA  30144}
\email{ted@avatar.kennesaw.edu}

\author{Michael~E.~Nord}
\affil{Department of Physics and Astronomy, University of New Mexico, 
Albuquerque \& Remote Sensing Division, Naval Research Laboratory,
Washington DC 20375-5351}
\email{Michael.Nord@nrl.navy.mil}

\author{T.~Joseph~W.~Lazio \& Namir~E.~Kassim}
\affil{Remote Sensing Division, Naval Research Laboratory,
Washington DC 20375-5351}
\email{Joseph.Lazio@nrl.navy.mil}
\email{Namir.Kassim@nrl.navy.mil}

\begin{abstract}
New high-resolution, wide-field 90~cm VLA\footnote{%
The National Radio Astronomy Observatory (NRAO) is a facility of the
National Science Foundation operated under a cooperative agreement by
Associated Universities, Inc.}
observations of the Galactic Center region by Nord et al.\ have
revealed twenty nonthermal filament (NTF) candidates.  We report 6~cm
polarization observations of six of these.  All of the candidates have
the expected NTF morphology, and two show extended polarization confirming
their identification as NTFs.  One of the new NTFs appears to be part
of a system of NTFs located in the \objectname[]{Sgr~B} region, 64~pc
in projection north of \objectname[]{Sgr~A}. These filaments cross the
Galactic plane with an orientation similar to the filaments in the
Galactic Center Radio Arc.  They extend the scale over which the NTF
phenomena is known to occur to almost 300~pc along the Galactic
plane. Another NTF was found in the Galactic plane south
of the \objectname[]{Sgr~C} filament but with an orientation
of~45\arcdeg\ to the Galactic plane.  This is only the second of~12
confirmed NTFs that is not oriented perpendicular to the
Galactic plane.  An additional candidate in the Sgr~C region was resolved
into multiple filamentary structures.  Polarization was detected only at the brightness peak of one of the filaments.  Several of these filaments run parallel to the Galactic plane and can be considered additional evidence for non-poloidal magnetic fields at the GC. Together the 90
and~6~cm observations indicate that the Galactic center magnetic field
may be more complex than a simple globally ordered dipole field.
\end{abstract}

\keywords {ISM:Galactic Center --- radio continuum}

\section{Introduction}\label{sec:intro}

Understanding the origin, evolution, topology and strength of the Galactic center (GC)
magnetic field(s) is a necessary prerequisite for constructing a
coherent picture of activity at the \hbox{GC}.  The discovery of the
\objectname[]{Galactic Center Radio Arc} (GCRA) \citep{y-zmc84}
demonstrated that magnetic structures with ordering on the scale of
$\sim 40$~pc exists in the \hbox{GC}.  The GCRA may also be part of an even a larger scale
structure known as the \objectname[]{Omega lobe} \citep{sh84,s85}, which is a
loop-like structure extending several hundred parsecs from the GCRA
and may connect with the Sgr~C filament at its other footpoint.
Ejection or expansion of magnetic fields from an accretion disk
\citep{uss85,hnp98} or a shocked Galactic wind \citep{c92,b-hc03} are
possibilities for the origin of these structures.  In addition, the
discovery of isolated nonthermal filaments throughout the inner few
hundred parsecs \citep[e.g.,][]{y-zb89,m94,lklh00} with orientations
largely perpendicular to the Galactic plane have been interpreted as
evidence for a space-filling poloidal field in the ionized medium
\citep{ms96,m98}.  A pervasive poloidal field could result from the
inflow of a primordial halo field toward the high concentration of
mass in the GC \citep{sf87,ccm00,c01} or from a nuclear star burst as
stellar winds and supernovae inject stellar fields into the
surrounding medium \citep{w02}.  Black hole accretion disks can also
generate magnetic fields \citep[e.g.,][]{crv94} that subsequently
spread throughout the galaxy \citep[and references
therein]{w02,dl90,k02}.  Thus, both ejection of dynamo generated
fields and inflow and concentration of primordial fields could be
acting independently to create the observed structures.  Although
these theoretical ideas are viable and reasonable they have not been developed
with sufficient rigor and detail to establish any definitive
conclusions on the nature of magnetism at the \hbox{GC}.

Observationally the nonthermal filaments (NTFs) are unique to the GC and are defined and identified by (1)~Extreme length to width
ratios (tens of parsecs long and only a few tenths of a parsec wide);
(2)~Nonthermal spectral indices; and (3)~High intrinsic polarization
\citep[for recent reviews see][]{bl01,lnlk03}.  Faraday rotation
measurements indicate that the NTF magnetic fields are aligned
longitudinally \citep[e.g.,][]{tihtksk86,y-zwp97,lme99}.  Of the nine
previously identified isolated NTFs---the most recent NTF discovery is
described by \cite{r03}---eight of them, along with the bundled NTFs in the
GCRA, are within 20\arcdeg\ to the perpendicular to the Galactic
plane.   From the distribution and lengths of NTFs, it has been 
suggested that the magnetic field in the ionized medium
is dominated by a poloidal field that extends many tens of parsecs
from the Galactic plane \citep[e.g.,][]{ms96,m98}.  [The lone NTF
parallel to the plane, assuming a GC distance of 8 Kpc \citep{r93}, is located southwest about~75~pc in projection
from the Galactic plane and the farthest from \objectname[]{Sgr~A}, approximately
225~pc in projection. It therefore could be in a region where the
field lines are merging with the toroidal field in the disk
\citep{laklg99}.]

Several estimates for the strength of NTF magnetic fields can be made.
Estimates based on radio continuum measurements and assuming
equipartition field strengths are of the order of a few tenths of a
milliGauss for several filaments (the \objectname[]{Northern Thread},
\citealt{lme99}; the \objectname[]{Sgr~C} filament,
\citealt{lklh00}; the \objectname[]{Snake filament},
\citealt{gnec95}).  The absence of any bending or distortion of the
NTFs against the strong ram pressure of the GC molecular clouds
indicates a field strength of about~1~mG (\citealt{y-zm87a,y-zm87b};
see \citealt{c01} for further discussion of this point).  Such a field
strength is considerably stronger than the general interstellar
magnetic field, which is typically no more than a few tens of
microGauss \citep[e.g.,][]{c99}.  The implied internal magnetic
pressures of the NTFs are so large that unless they are confined by a
comparable pressure, their lateral dimensions would
expand well beyond the observed few tenths of a parsec on timescales
short compared to their synchrotron lifetime.  Confinement is not an issue if there is a 
$\sim 1$~mG field filling the entire region.  In that case the NTFs are those
flux tubes that happen to be illuminated with high energy electrons
that are energized by some local interaction.  The energy density of
such a space-filling field is quite high, of the order of $4 \times
10^{-8}$~ergs~cm${}^{-3}$ and the total magnetic energy is $\sim 6 \times
10^{54} (R/75)^2(L/300)$~ergs, where $R$ and $L$ are the radius and length
(in pc) of the cylindrical region over which the NTF phenomenon is
found.  

However, alternatives to the pervasive field interpretation are also viable (e.g., Yusef$-$Zadeh 2003; \cite{sl99}). In the cometary model of \cite{sl99}, the NTFs are magnetized wakes generated by the interaction of molecular clouds with a GC wind.  Considerable evidence for a Galactic wind is derived from X-ray
observations of hot expanding gas \citep{kmstty96}.  The existence of
a GC wind and related outflows is discussed in the
context of recent high resolution infrared observations by
\cite{b-hc03}.  In the cometary model the wind is assumed to advect a
weak $\sim 10$~$\mu$G magnetic field that is amplified by a factor of
roughly 100 as the field is stretched by the flow and wraps around the
obstacle, a molecular cloud, forming an elongated magnetotail
\citep[for simulations, see][]{gmrj00,dels02}).  In this
scenario the total required magnetic energy is reduced by several orders of
magnitude and presumably NTFs with every orientation could be observed
since they are viewed in projection.  

The above discussion indicates that additional observations are required to
to establish a consensus interpretation of the NTF phenomenon and it's relationship to galactic center magnetism.  
This paper reports new total and polarized intensity observations of a
number of NTF candidates that challenge the interpretation of a
space-filling globally ordered poloidal field.  In \S\ref{sec:observe}
we describe the observations and analysis and in \S\ref{sec:discuss}
we discuss the implications of our observations.

\section{Observations and Results}\label{sec:observe}

\begin{figure}
\plotone{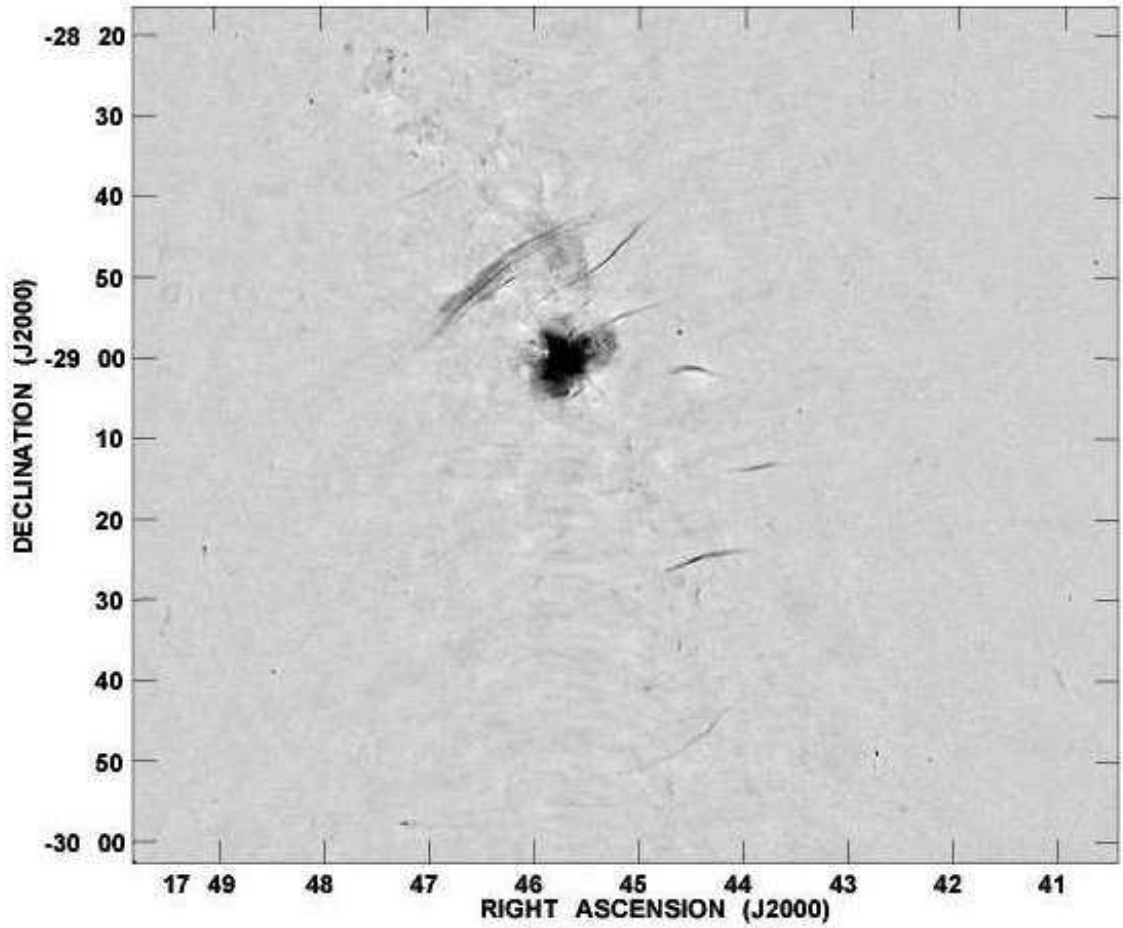}
\caption[]{The inner $0.8\arcdeg \times 1.0\arcdeg$ of the Galactic
center region at~90~cm \citep{nbhlklad03,nlkhld03}.  The resolution is $7\arcsec \times 12\arcsec$.  This image was generated using a non$-$linear transfer function to simultaneoulsy show the detail in the Sgr A region and the fainter NTFs}
\label{fig:fig1.eps}
\end{figure}

Recently improved wide-field, high sensitivity, high resolution, VLA
imaging of the GC at 90~cm \citep{nbhlklad03,nlkhld03} has revealed
twenty new NTF candidates, identified solely on the basis of their
morphology.  Figure~\ref{fig:fig1.eps} shows the inner $0.8\arcdeg
\times 1.0\arcdeg$ of the full 90~cm image produced by
\cite{nlkhld03}.  It was constructed from a combination of VLA A- and
B-configuration observations, with a resulting resolution of
$12\arcsec \times 7\arcsec$.  Although it is not sensitive to
large-scale, low surface brightness features, this image shows all
previously known NTFs except for \objectname[NTF]{NTF~359.44$+$0.39}
which was significantly resolved.  The NTFs appear because their angular extent in one
dimension is comparable to or smaller than the beam.

The previously known NTFs are mainly perpendicular to the Galactic
plane.  However, the orientations of the candidate NTFs are more
diverse, with several nearly parallel to the plane.  The candidates
are also considerably shorter, with many less than 10~pc in length.
The surface brightnesses of the candidates (15--20~\mjybm) are roughly
a factor of~4 or so less than the more prominent NTFs
\citep{nbhlklad03,nlkhld03}.  In retrospect a number of these
candidates have been detected at higher frequencies.  The candidate
NTFs in the \objectname[]{Sgr~A} region can be seen on the high
resolution 20~cm image of \cite{lme99}, who referred to shorter
filamentary structures as ``streaks.''  In addition several candidates
near \objectname[]{Sgr~C} were detected by \cite{ls95} at~18~cm.

Although the morphologies of these NTF candidates are suggestive,
spectra could not be constructed for all of them nor had any
polarization observations been conducted.  In order to confirm their
status as NTFs, we conducted VLA observations of six (6)  of these
sources at~6~cm in~2002 October.  The observations were conducted with
the VLA in the CnB configuration, providing a resolution of $4\arcsec
\times 3\arcsec$ (significantly higher than at~90~cm).  Full
polarization information was recorded.

The visibility data were calibrated and imaged using standard
techniques within \textsc{aips}.  Both total intensity (Stokes~I) and
linearly polarized intensity (Stokes~Q and~U) images were formed.  The
Galactic nonthermal background along the plane remains significant
even at~6~cm and, despite integration times of several hours per
source, it was not possible to achieve signal-to-noise ratios
exceeding 5 in the total intensity images.  The low signal-to-noise
ratio also compromised our ability to detect polarization.  In some
cases, we were able to detect the total intensity emission from a
source but no polarized emission.  For the total linear polarized
intensity ($L \equiv \sqrt{Q^2 + U^2}$), a Rice-Nakagami distribution
is obtained, and the upper limits we quote are relative to the rms
noise level in the Q and~U images, $\sigma_{Q,U}$.  Nonetheless, these
various limitations did not preclude our identifying a number of these
candidates as NTFs.

\begin{figure}
\includegraphics[angle=-90,scale=0.7]{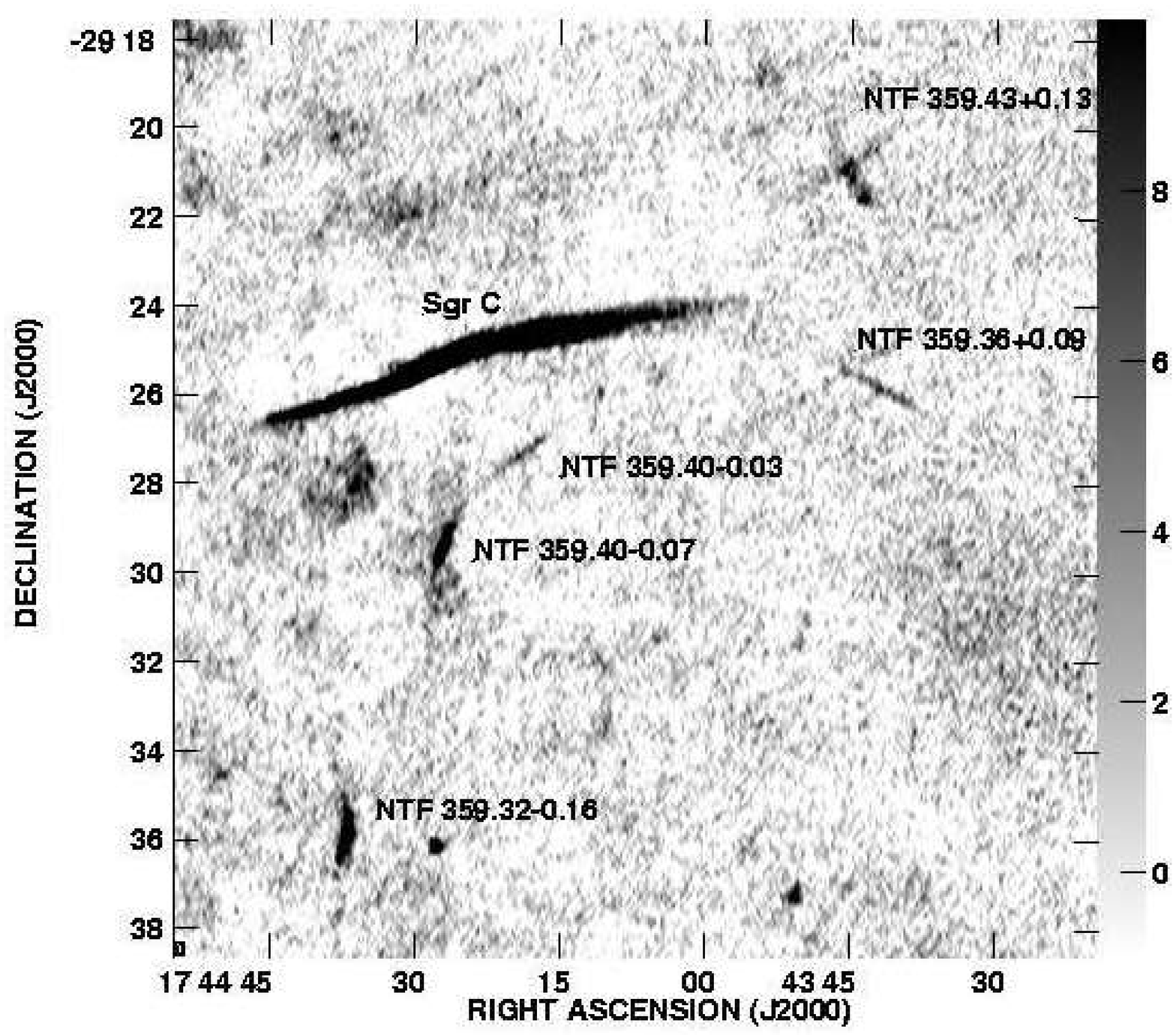}
\caption[]{The \protect\objectname[]{Sgr~C} region at~90~cm.
Candidate NTFs are labeled with Galactic coordinates.  The grey scale is linear between~$-$1 and~10~mJy~beam${}^{-1}$.}
\label{fig:fig2.eps}
\end{figure}

\begin{figure}
\plotone{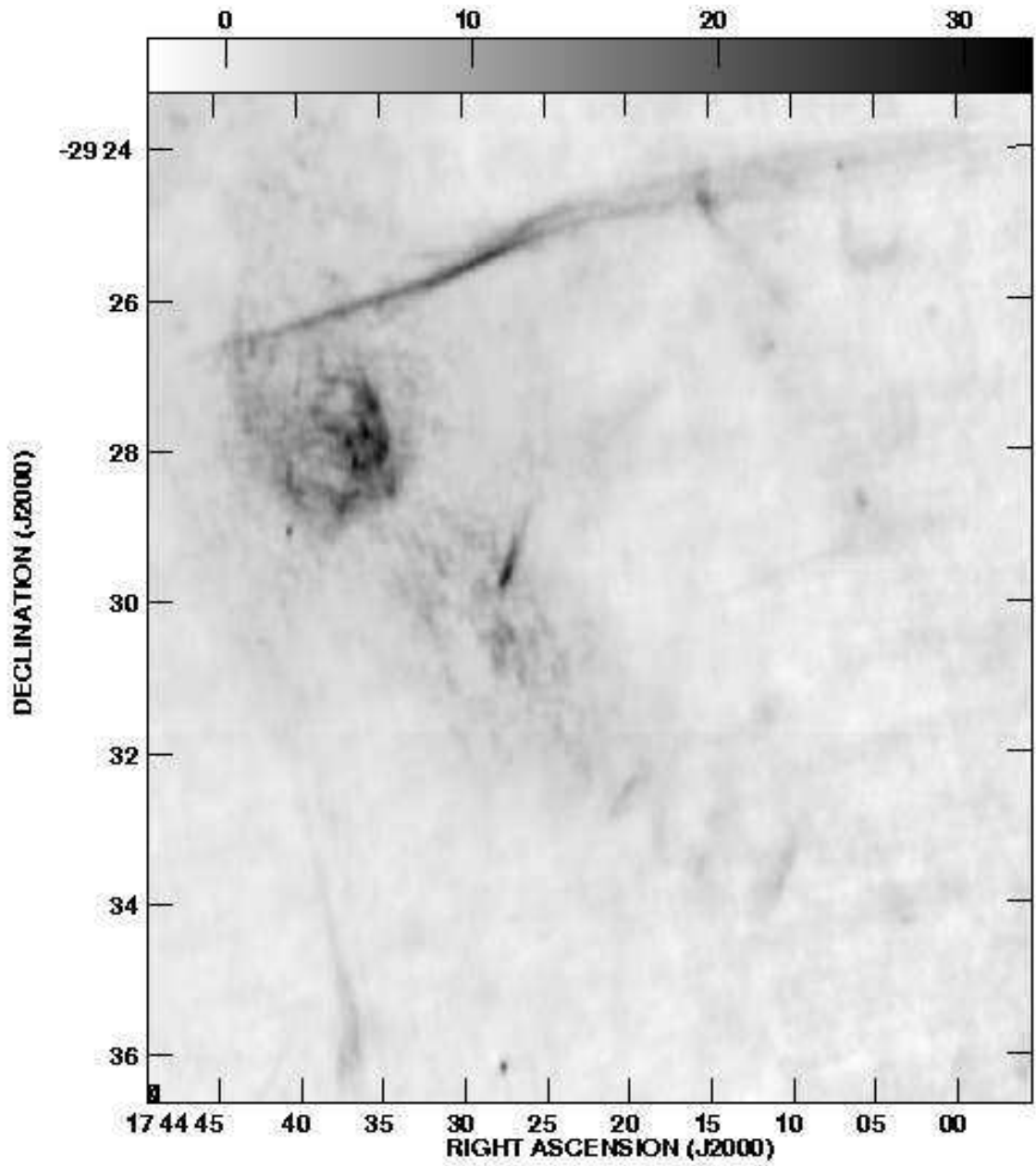}
\caption{The \protect\objectname[]{Sgr~C} region at~18~cm from
\cite{ls95} showing \protect\objectname[]{G359.32$-$0.16} directly south of the H II region at 17 44 37 and -29 36.  The resolution is $7.5\arcsec \times 4\arcsec$.}
\label{fig:fig3.eps}
\end{figure}

\begin{figure}
\plotone{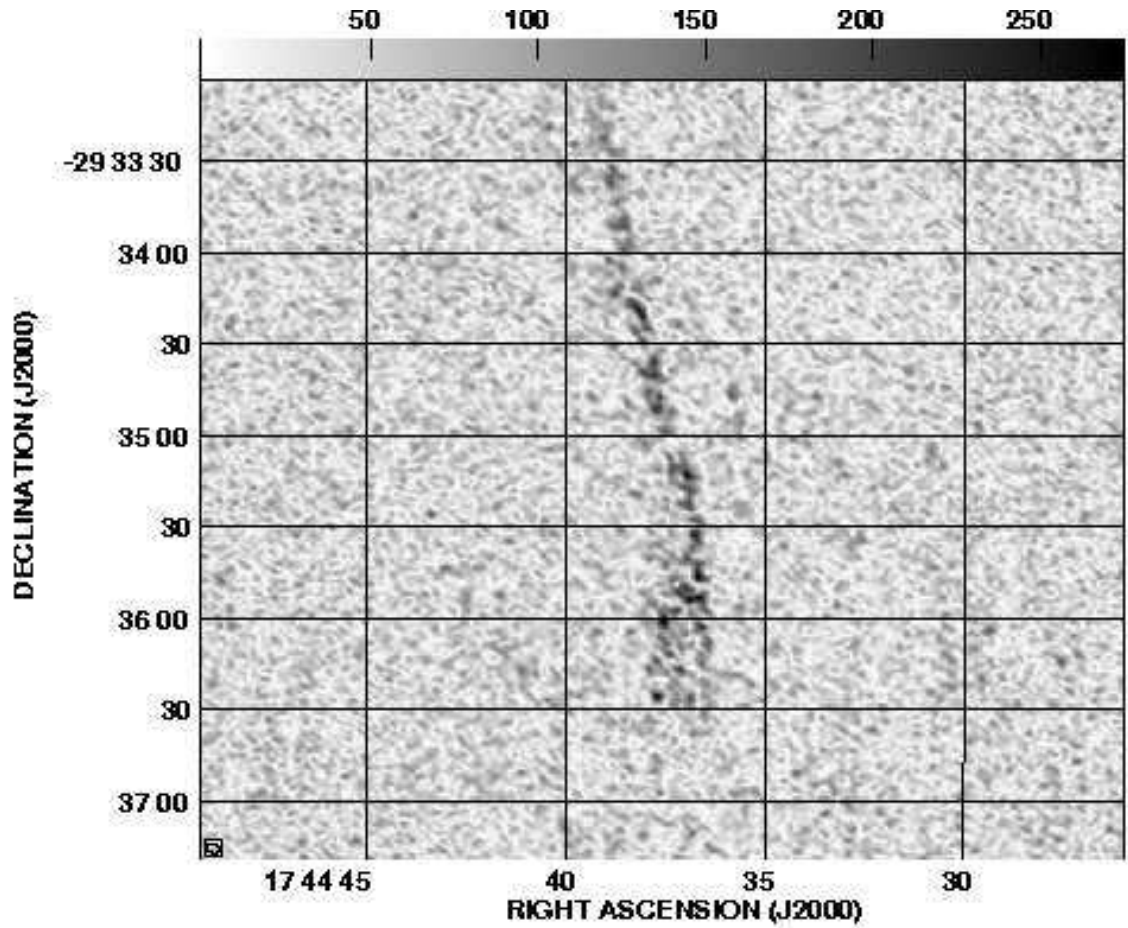}
\caption[]{The polarized intensity image of
\protect\objectname[NTF]{NTF~359.32$-$0.16} at~6~cm.  The resolution
is $3.75\arcsec \times 3\arcsec$, and the rms noise level is 25~$\mu$Jy~beam${}^{-1}$.  The gray scale is linear between~0 and~274~$\mu$Jy~beam${}^{-1}$.}
\label{fig:fig4.eps}
\end{figure}

\begin{figure}
\epsscale{0.75} 
\plotone{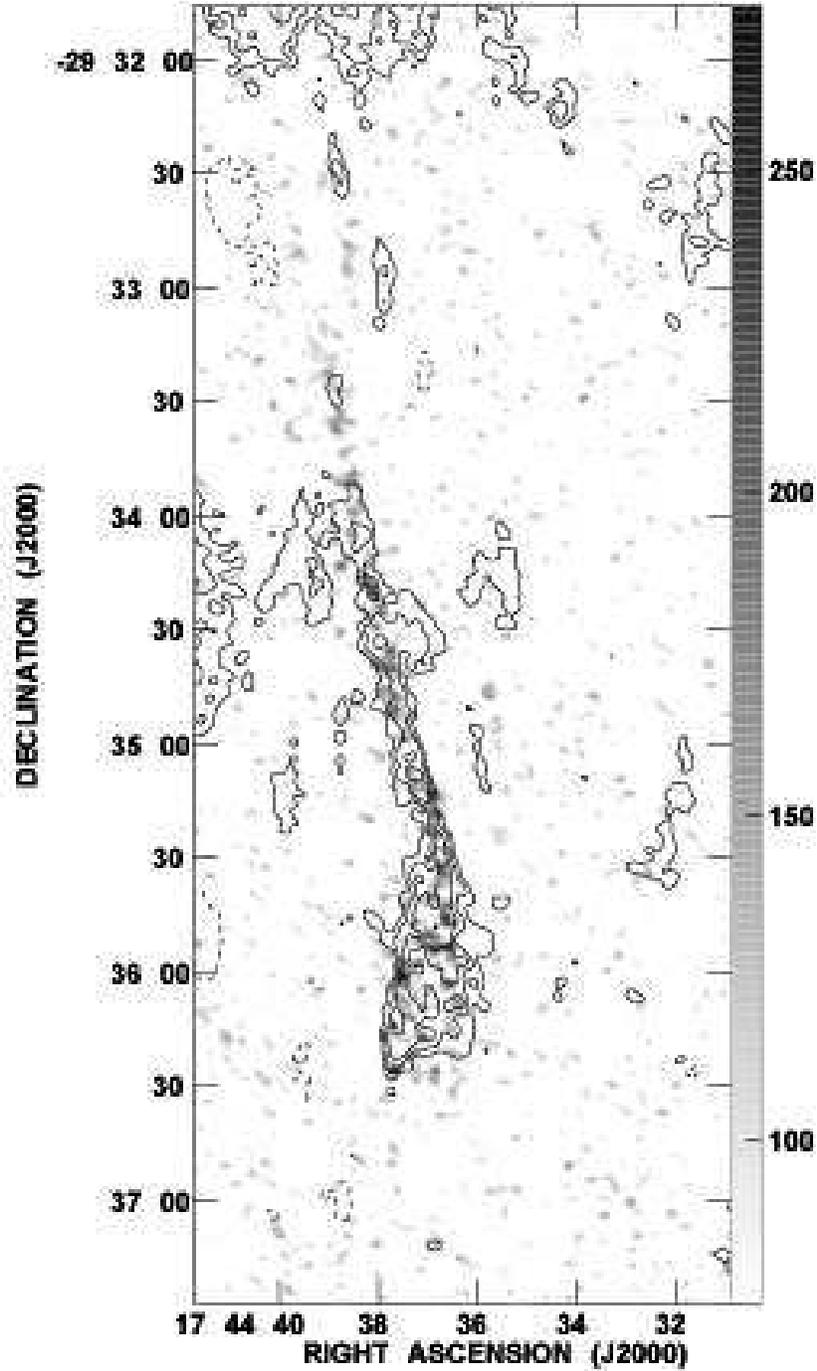}
\caption[]{Total and polarized intensity images of
\protect\objectname[NTF]{NTF~359.32$-$0.16} at~6~cm. The beam is $3.75\arcsec
\times 3\arcsec$.  The contours show the total intensity and are
0.2~\mjybm $\times$ $-2$, 1.25, 2, 3, 5, 7.07, and~10, with the rms
noise level in the total intensity image being 0.2~\mjybm.  The gray
scale shows the polarized intensity and is linear between~75
and~275~$\mu$Jy~beam${}^{-1}$.}
\label{fig:fig5.eps}
\end{figure}

\begin{figure}
\plotone{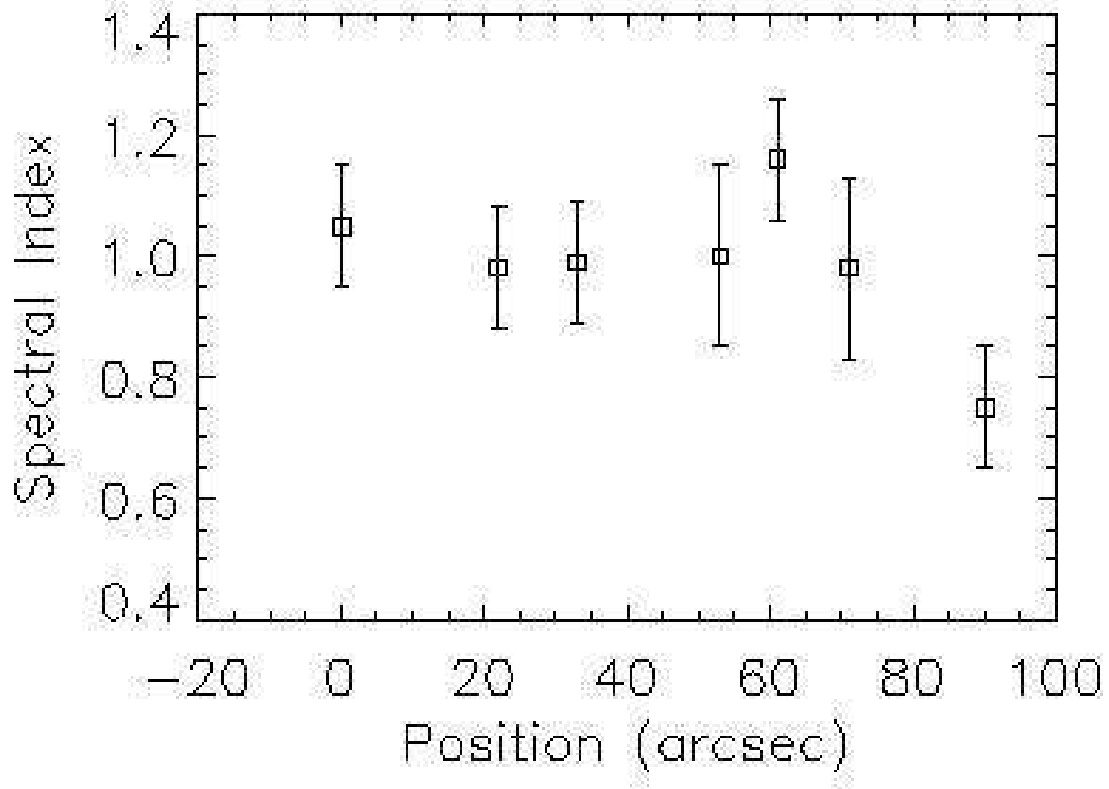}
\caption[]{The 6/18~cm spectral index as a function of length along
\protect\objectname[NTF]{NTF~359.32$-$0.16}.  The error bars were determined from uncertainties in the baselines of the individual cross$-$cuts. The 0 position is located at 17 44 37.5, -29 36 01.  The last position is located at 17 44 37.5, -29 34 31.}
\label{fig:fig6.eps}
\end{figure}

\begin{figure}
\plotone{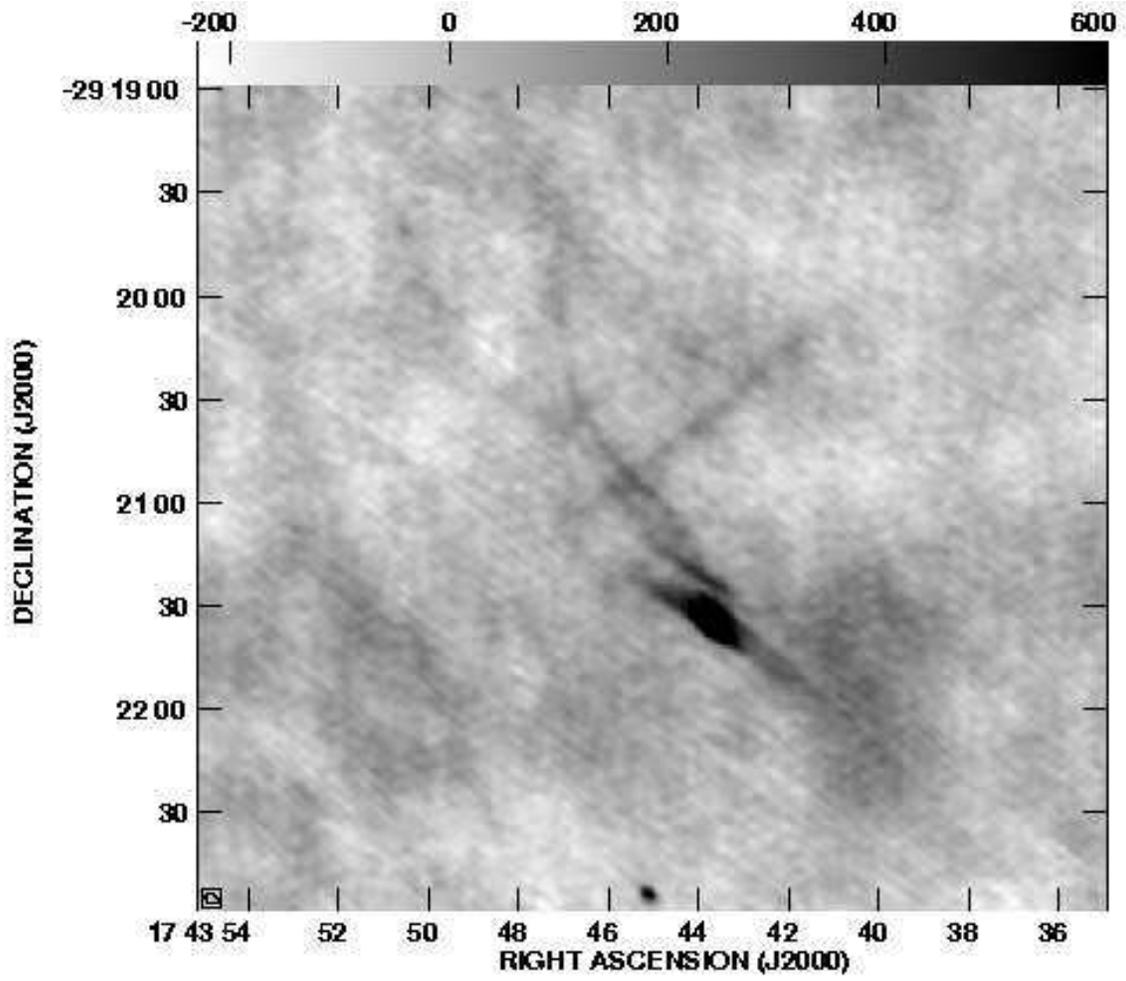}
\caption[]{The total intensity image of
\protect\objectname[NTF]{NTF~359.43$+$0.13} at~6~cm.  The resolution
is $3.9\arcsec \times 2.8\arcsec$, and the rms noise level is 66~$\mu$Jy~beam${}^{-1}$.  The gray scale is linear between~$-$228 and~600~$\mu$Jy~beam${}^{-1}$.}
\label{fig:fig7.eps}
\end{figure}

\begin{figure}
\begin{center}
\includegraphics[angle=-90,scale=0.7]{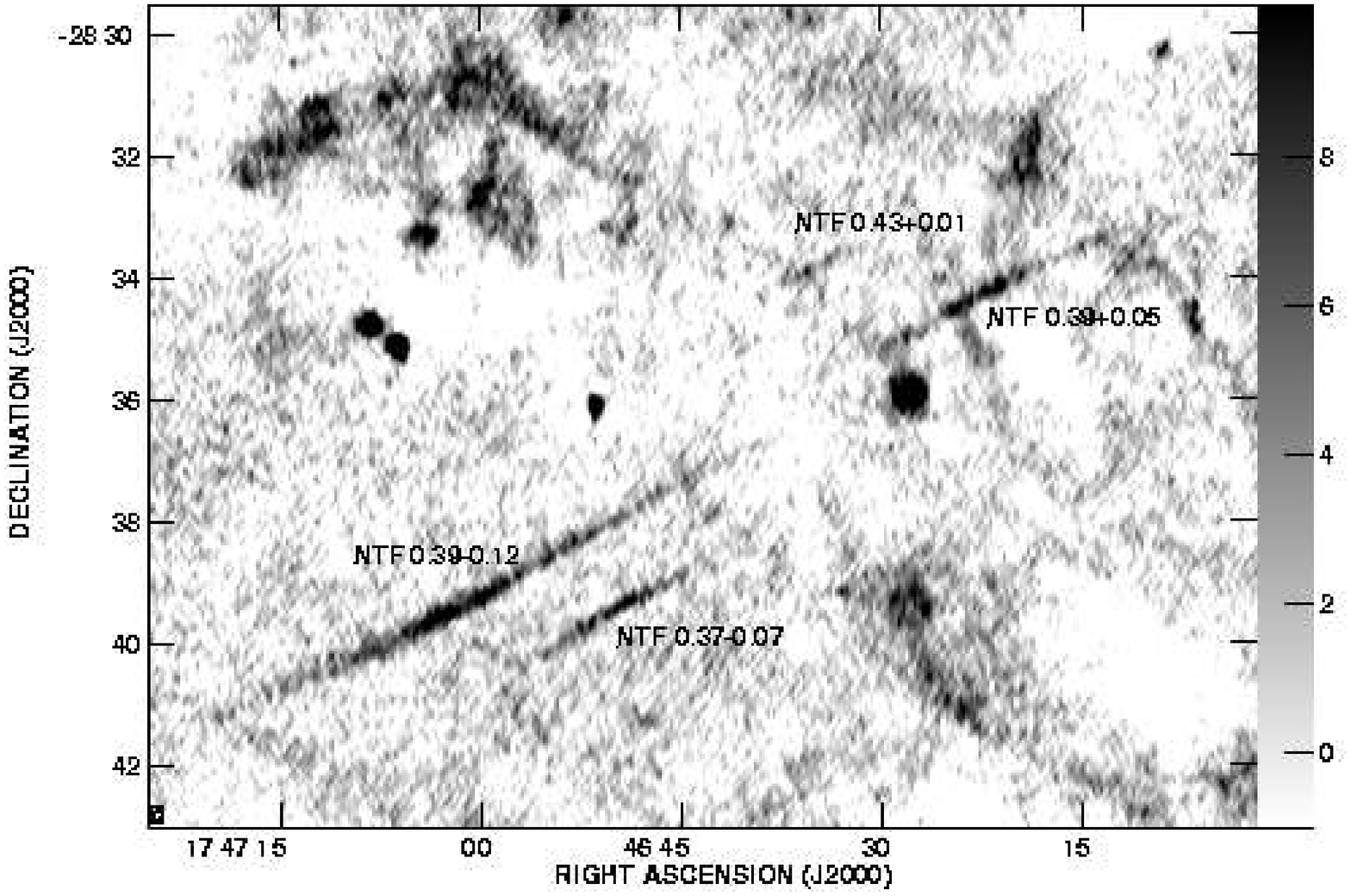}
\end{center}
\caption[]{The \protect\objectname[]{Sgr~B} region at~90~cm. Candidate
NTFs are labeled with Galactic coordinates.  The grey scale is linear between~$-$1 and~10~mJy~beam${}^{-1}$.}
\label{fig:fig8.eps}
\end{figure}

\begin{figure}
\begin{center}
\includegraphics[angle=-90,scale=0.7]{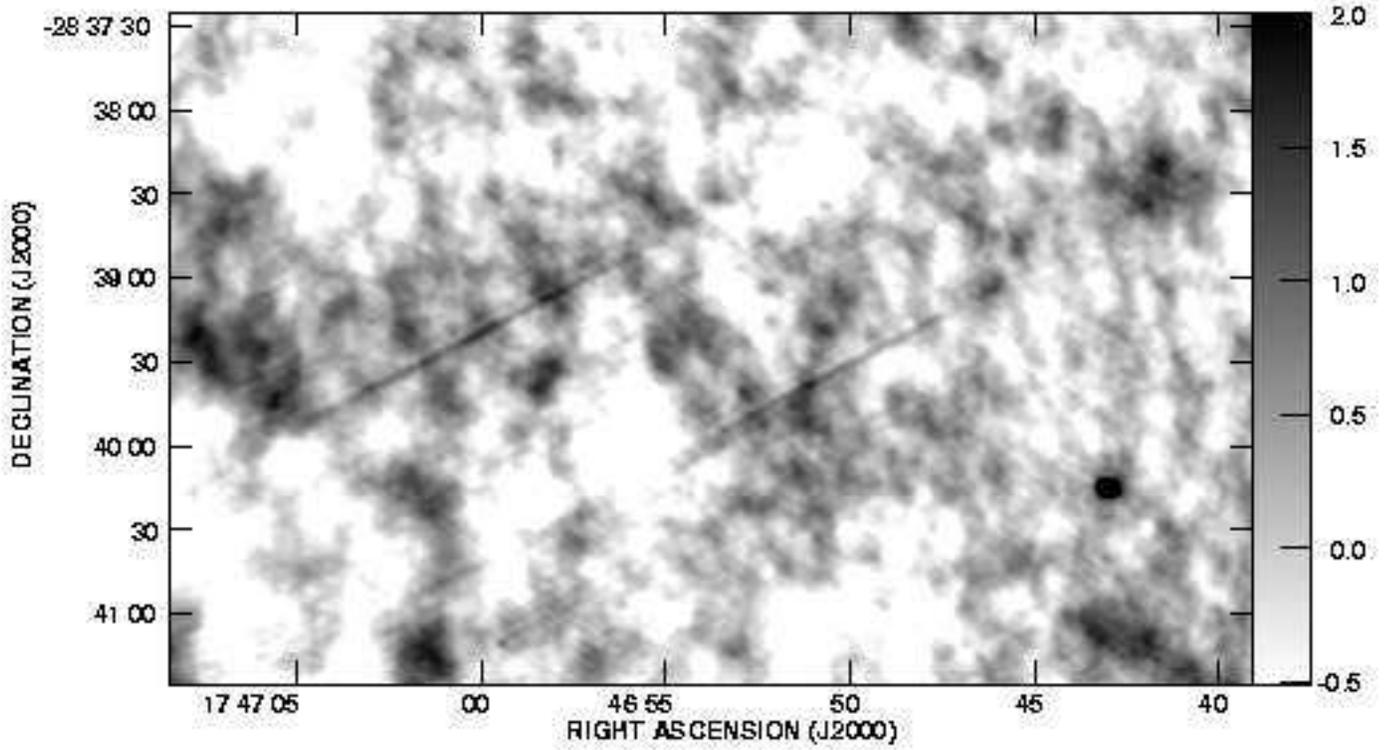}
\end{center}
\caption[]{The newly discovered filaments in the \protect\objectname[]{Sgr~B}
region at~6~cm.  The straight filamentary nature of
\protect\objectname[NTF]{NTF~0.37$-$0.07} and
\protect\objectname[NTF]{NTF~0.39$-$0.12} is in this total
intensity image.  As shown in the next figure \protect\objectname[NTF]{NTF~0.39$-$0.12} is highly polarized.  \protect\objectname[NTF]{NTF~0.39$+$0.05} is too faint to be seen at this wavelength.  The gray scale is linear between~$-0.5$ and~2~\mjybm.  The beam is $3.6\arcsec \times 2.9\arcsec$, and the rms noise level is 0.64~\mjybm.}
\label{fig:fig9.eps}
\end{figure}

\begin{figure}
\begin{center}
\includegraphics[angle=-90,scale=0.7]{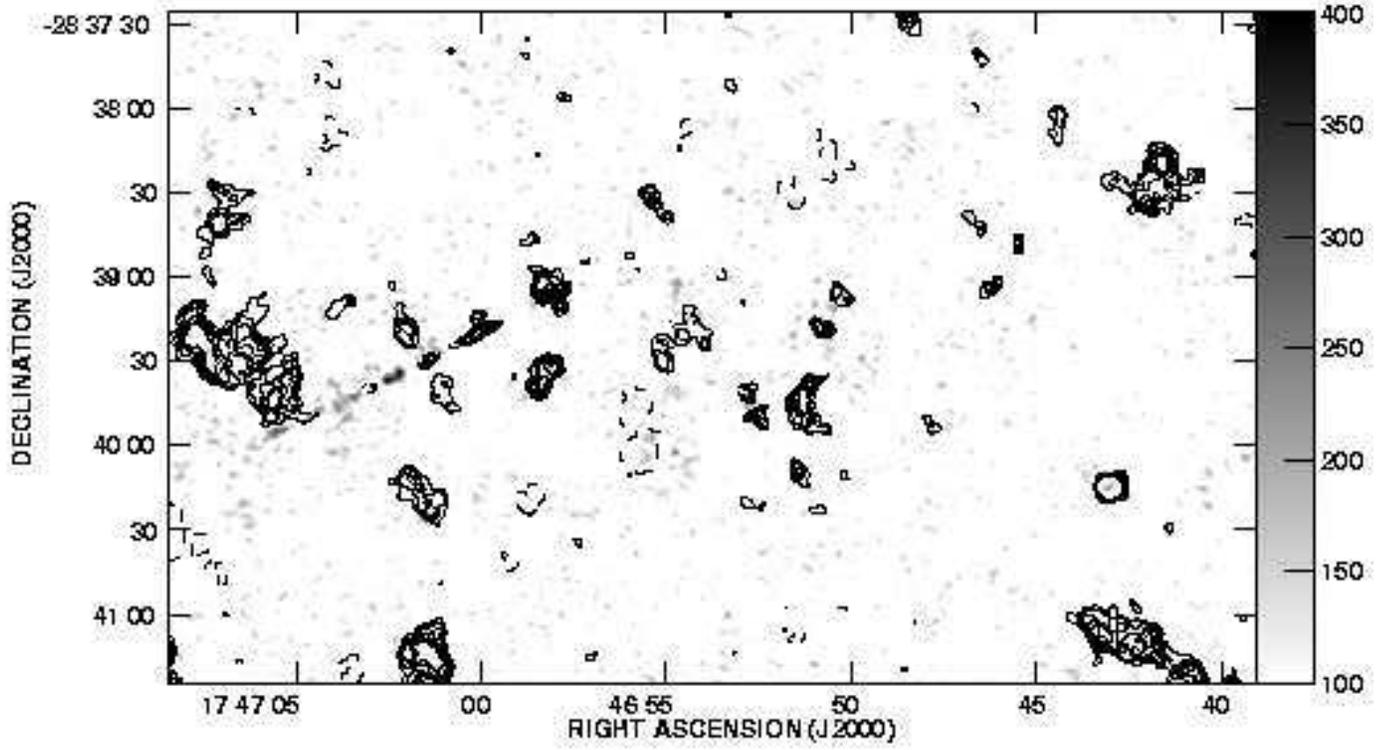}
\end{center}
\caption[]{The total and linearly polarized intensity of \protect\objectname[NTF]{NTF~0.39$-$0.12}.  The contours show the total intensity, and are 0.64~\mjybm\ $\times$ $-2$, 1.5, 1.75, 2, 2.25, and~2.5.  The gray scale shows the linearly polarized
intensity and is linear between~0.1 and~0.4~\mjybm.}
\label{fig:fig10.eps}
\end{figure}

We concentrated on several candidates in the \objectname[]{Sgr~B} and
\objectname[]{Sgr~C} regions, which we describe individually.
\begin{description}
\item[NTF~359.32$-$0.16]
Figure~\ref{fig:fig2.eps} shows a portion of the 90~cm image in the
region of \objectname[]{Sgr~C}. \objectname[NTF]{NTF~359.32$-$0.16} is
located due south of the \objectname[]{Sgr~C} \ion{H}{2} region. This
source is also detected at~18~cm \citep{ls95}, as
Figure~\ref{fig:fig3.eps} shows.  The surface brightness at~18~cm is not
uniform.  The peak brightness ($\approx 7$~\mjybm) occurs at the
southern terminus where the source is clearly wider, and the
brightness decreases uniformly to~1~\mjybm\ at the northern end.
Intensity profiles perpendicular to the source's major axis show a
clear double peak at the south end, suggesting there may be two
filaments in this system.  The length of this source, assuming a
Galactic center distance of~8~Kpc is $\sim 8$~pc. The 90~cm resolution
is a factor of~3 poorer and does not show any structural details.

Figure~\ref{fig:fig4.eps} shows the 6~cm polarized
image of \objectname[NTF]{NTF~359.32$-$0.16}.  Figure~\ref{fig:fig5.eps} shows the polarized intensity in grey scale overlain with contours of total intensity.  The polarization is patchy, similar to what is observed for other NTFs.  The peak fractional polarization is 65\% and the average fractional polarization is $\sim$50\%.  This is consistent with the polarizations seen in other NTFs.  The
polarized image has a significantly better signal-to-noise ratio
than the total intensity image, which is limited by the strong
background emission from the Galactic plane.  The object is also 8~pc
long at~6~cm and $\leq$ 0.5~pc wide.  Given the similar beam sizes
at~6 and~18~cm, we convolved the 6~cm image to the resolution of the
18~cm image and formed a series of intensity profiles at constant
declination.  These profiles were used to determine the peak
brightness as a function of length along the structure and estimate
the spectral index.  The different spatial frequency ($u$-$v$)
coverage in the~6 and~18~cm observations complicate the determination
of an absolute spectral index.  However, this analysis should reveal
relative changes of spectral index with position.  As
Figure~\ref{fig:fig6.eps} illustrates, to within the uncertainties, the
spectral index is constant, $\alpha_{6/18} \sim -1$ ($S \propto
\nu^{\alpha}$), along the length of the filament.  This constancy with position is 
consistent with other well-studied NTFs \citep{lme99,lklh00}.  Spectral curvature appears to be another characteristic of the \hbox{NTF} phenomenon.  Typical 20/90~cm and 6/20~cm spectral indices are between -0.4 and -0.6 but above 5 GHz the spectrum turns over \cite{lme99}.  Using the same method as described above we estimated the 18/90~cm spectral index near the location of the peak flux.  We find $\alpha_{18/90} \sim -0.1$.  Thus our 6/18~cm index is steeper than typical but our 18/90~cm index is flatter.  As we mentioned determining an absolute spectral index is problematic but we can confidently 
conclude that this source is nonthermal.  Based on morphology, percentage polarization, and the nonthermal spectral index we classify this
source as an \hbox{NTF}.

\item[NTF~359.43$+$0.13]
This candidate lies northwest of the \objectname[]{Sgr~C} filament.
At~90~cm it has a distinctive \texttt{X}-shape.
Figure~\ref{fig:fig7.eps} shows that at~6~cm the \texttt{X} is
resolved into a very bright, slightly elongated source; two, or
perhaps three, longer filaments with significant curvature running
roughly parallel to the Galactic plane; and another filament
perpendicular to these.  Polarization of $\sim 10\%$ was detected near the location of the peak brightness.  The signal-to-noise ratio along the
filamentary structures is only 2--3. Unfortunately, at this low
signal-to-noise ratio, we would not have detected polarized emission,
even if these filaments were NTFs with high polarizations.  Thus we can not state categorically that these sources are NTFs.  However given their morphology it would be difficult to build a case for an alternative classification. 

\item[NTF~359.40$-$0.07]
This source is the brightest NTF candidate at 90~cm and lies 5~pc
south in projection from the \objectname[]{Sgr~C} filament.  It was also
detected at 18~cm \citep{ls95}.  We detect \objectname[NTF]{NTF~359.40$-$0.07} in total intensity at 6~cm, but we do not detect what appears to be a faint extension of this source in the 90~cm image (\objectname[NTF]{NTF~359.40$-$0.03}).  Using the
90~cm image to delineate the approximate extent of the source, we find
the total linear polarization intensity distribution within this
region to be noise-like with no linear polarized emission
above~4.5$\sigma_{Q,U}$, where $\sigma_{Q,U} = 20$~$\mu$Jy~beam${}^{-1}$.  The lack of polarized emission (if the source is in fact polarized!) may be a consequence of the strong background emission.  To obtain a rough estimate of the spectral index we convolved the 6~cm map to 18~cm map and made a few cross$-$cuts near the location of the peak emission.  We find that the 6/18~cm spectral index is $\alpha_{6/18} \sim -0.1$.  Employing a similar procedure for the 18 and 90~cm maps we estimate that the 18/90~cm spectral index is $\sim 0.1$.  There is considerable error in convolving the higher resolution maps to a larger beam in a confused region such as this and we hestitate to draw any firm conclusions from these results.  Further observations will be required to classify this source.

\item[NTF~0.37$-$0.07, NTF~0.39$+$0.05, and NTF~0.39$-$0.12]
These three linear features are in the \objectname[]{Sgr~B} region.
Figure~\ref{fig:fig8.eps} shows the region at~90~cm
\citep{nbhlklad03,nlkhld03}, Figure~\ref{fig:fig9.eps} shows the
region at~6~cm total intensity, and Figure~\ref{fig:fig10.eps} shows~6~cm total and linear polarized intensity overlayed.  Due to the low signal-to-noise ratio only
\objectname[NTF]{NTF~0.39$-$0.12}, was detected in polarization.  The peak fractional polarization is 45\% while the average fractional polarization 
is $\sim$35\%.  It appears that \objectname[NTF]{NTF~0.39$-$0.12}
and \objectname[NTF]{NTF~0.39$+$0.05} could be part of the same
filament.  Their clear linear morphology and similar orientation to
the filaments in the GCRA strongly suggests they are NTFs.

They are located about~64~pc in projection from \objectname[]{Sgr~A}
and are the first filaments found  north of the \hbox{GCRA}.  Although
they are parallel to the GCRA filaments, they are located farther
south of the Galactic plane.  Thus, these objects extend the length
scale (and therefore volume) over which the NTF phenomenon is know to
occur to almost 300~pc along the Galactic plane.  If the NTFs are a
consequence of a global magnetic field, the field must be organized on
at least this size scale.
\end{description}

\section{Discussion \& Conclusions}\label{sec:discuss}

The 6~cm observations show that all the observed NTF candidates are
filamentary in nature, with a several of them showing polarization.  Although only 3 of the 6 candidates exhibited the expected polarization, the low
signal-to-noise would have precluded detecting polarized emission from
the remainder, even if they were polarized at the level typical of
NTFs.  Generalizing these results, we conclude that many of the 90~cm
NTF candidates probably are true NTFs.  Consequently there may be a
large population of shorter, low surface brightness NTFs.

Are these shorter, low surface brightness NTFs the same class of
phenomenon as the longer, brighter NTFs that are already known? 
\cite{nbhlklad03} shows that the typical surface brightness of the new
filaments is approximately 20~\mjybm\ at~90~cm; the surface brightness
of the previously known filaments is 80~\mjybm.  The flux density
$S_\nu$ as a function of frequency~$\nu$ from a synchrotron source of
volume~$V$ is $S_\nu \propto N_0 VB^{-\alpha+1} \nu^{\alpha}$ where
$N_0$ is a number density of synchrotron-emitting electrons and~$B$ is
the magnetic field within the volume.  The widths of the new filaments
are comparable to those of the previously known, prominent NTFs, of
order a few tenths of a parsec.  Assuming that these are approximately
cylindrical objects, then the volume element per unit length is the
same for both the previously known and newly identified filaments.
For a spectral index $\alpha \approx -1$, the luminosity of a
synchrotron source scales with magnetic field as $B^2$.  Therefore, in
order to reproduce the observed range in surface brightnesses requires
that the product $N_0B^2$ vary by only a factor of~4.  Either a factor
of~4 variation in the synchrotron-emitting electron number density or
a factor of~2 in the magnetic field strength is sufficient.  Of
course, if both $N_0$ and~$B$ vary, even smaller variations in each
are required in order to obtain the required variation in $N_0B^2$.
We conclude that the same physical process could be responsible for
the entire population with small local variations in magnetic field
strength and/or energetic particle density accounting for any
differences.

A larger number of NTFs offers additional insights into the structure of any large$-$scale magnetic field.  \objectname[NTF]{NTF~359.32$-$0.16} is oriented at an angle of~45\arcdeg\ to the Galactic plane and is only the second NTF that is not oriented more-or-less perpendicularly with respect to the Galactic plane.  It is also the considerably closer to \objectname[]{Sgr~A} (87~pc in projection) than the other parallel NTF, \objectname[NTF]{NTF~358.85$+$0.47} (the
Pelican).  There is considerable evidence that the magnetic field in
the neutral medium along the Galactic plane in the \objectname[]{Sgr~A} and \objectname[]{Sgr~B} regions is parallel to the plane \citep[e.g.,][]{novaketal03}\footnote{%
Additional results on the GC magnetic field are derived from submillimeter and far infrared observations of the neutral medium.  The polarized infrared emission from dust grains embedded within the molecular clouds along the Galactic plane suggest 
that the magnetic field is parallel to Galactic plane and therefore
toroidal \citep{novaketal03}.  Estimates for the strength of this
field also indicate a few mG \citep{cnhdvdd03}.} suggesting that there is a large-scale toroidal field in the neutral medium.  Given its
45\arcdeg\ orientation, it seems unlikely that this NTF is related to
a toroidal field.  Furthermore SCUBA images \citep{pierce-priceetal00}
of the Galactic center indicate that the thermal emission from the
\objectname[]{Sgr~C} environment is much less than in the
\objectname[]{Sgr~A} or \objectname[]{Sgr~B} regions.  Thus, the
region near \objectname[]{Sgr~C}, including
\objectname[NTF]{NTF~359.32$-$0.16} may not be dominated by the
neutral medium.  

\begin{figure}
\plotone{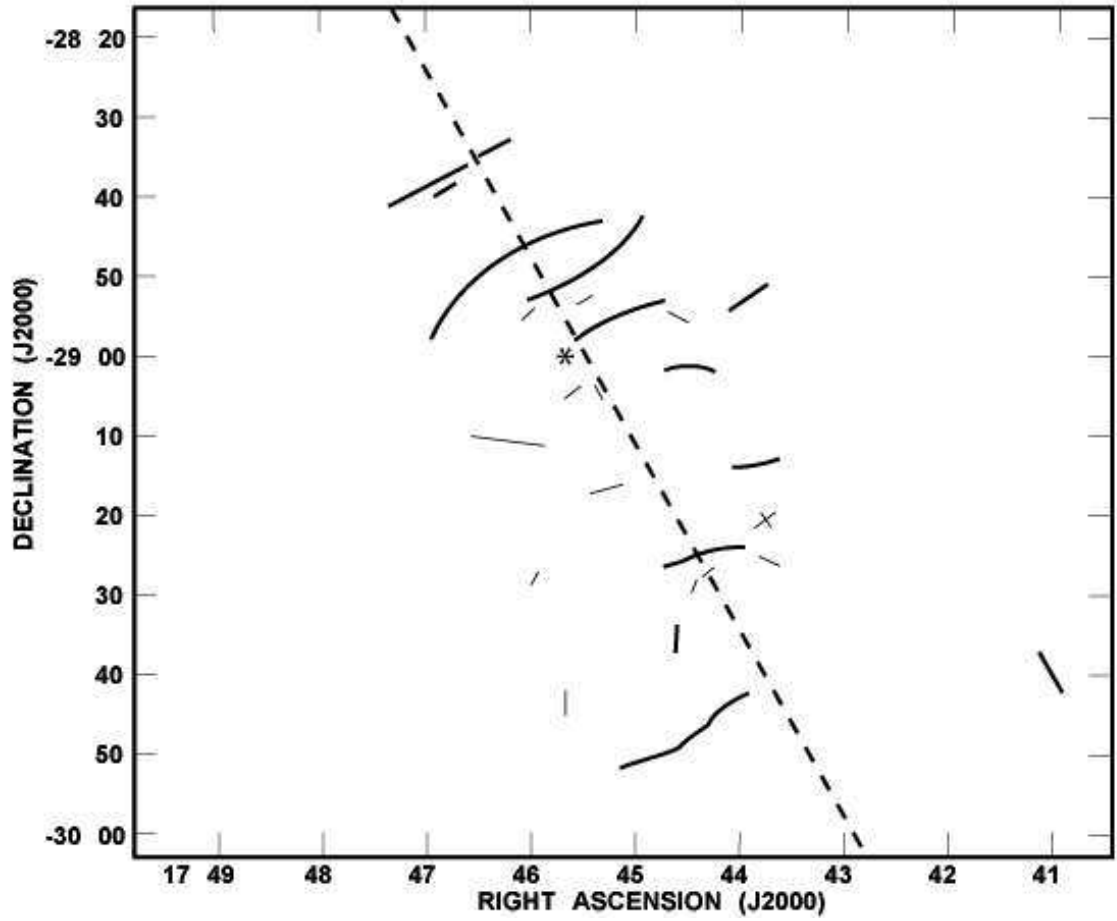}
\caption[]{Schematic of the GC region based on the 90~cm wide$-$field image.  Confirmed NTFs are shown in heavy dark lines.  NTF candidates are shown in thin lines. The Galactic plane is the dashed line and Sgr A$^*$ is marked with an asterisk.}
\label{fig:fig11.eps}
\end{figure}

Figure~\ref{fig:fig11.eps} is schematic of the GC region based on the wide$-$field image of Nord et al (2003a,b).  The confirmed NTFs (including those from this work) are shown in the heavy dark lines while the NTF candidates are denoted by the thin lines.  Confirmation of additional candidates would definitively rule out a global dipole field model.  However, the present data is already difficult to reconcile with global ordering.   Figure 10 shows that there are
several cases where two NTFs, which are close together in projection,
have very different orientations.  In order to preserve $\nabla \cdot
B =0$ in a region of space, distinct field lines must be more or less
parallel on scales smaller than any gradient scale.  The northern end
of \objectname[NTF]{NTF~359.32$-$0.16} is 15~pc in projection from
the \objectname[]{Sgr~C} filament and has an orientation $\sim
45\arcdeg$ to it.  The \objectname[]{Sgr~C} filament is about~30~pc
long and, although it has a nonuniform brightness, it exhibits a
constant spectral index with length, suggesting that it is tracing a
uniform magnetic field.  For a dipole magnetic field $B/\nabla B =
r/3$, where $r$ is the radial distance from the dipole.  For a dipolar
field centered on \objectname[]{Sgr~A} at the location of
\objectname[NTF]{NTF~359.32$-$0.16} (87~pc in projection) the gradient
scale length is almost 30~pc, consistent with the \objectname[]{Sgr~C}
\hbox{NTF}. However, if \objectname[NTF]{NTF~359.32$-$0.16} is at the
same distance with a $45\arcdeg$ difference in orientation, it could
not be tracing the same global, divergence-free field.  Another
example is the northern thread \objectname[NTF]{NTF~0.08$+$0.15} and
the NTF candidate \objectname[]{G359.90$+$0.19}.  These two are
separated by only about~5~pc in projection and both are about~30~pc in
projection from \objectname[]{Sgr~A}, yet their orientations differ
by~25\arcdeg.

These differences in orientation cannot be explained as due to
differences in their relative distances.  While little is known about
their relative separations, as noted earlier the NTF phenomenon occurs
over a linear distance of roughly 300~pc along the Galactic plane
($-1\arcdeg \lesssim \ell \lesssim 0\fdg4$).  It is reasonable to
assume that the NTFs are also distributed over a comparable distance
along the line of sight.  If this is the case, large apparent changes
in orientation could be obtained in a dipolar field by filaments well
above the Galactic plane and located at different distances along the
line of sight (e.g., the northern thread
\objectname[NTF]{NTF~0.08$+$0.15} and the NTF candidate
\objectname[]{G359.90$+$0.19}).  Even so, if the GC magnetic field is
poloidal and dominated by a dipolar component, all filaments crossing
the Galactic plane should be nearly perpendicular to the Galactic
plane, regardless of their distance along the line of sight.  The
different orientations of the \objectname[]{Sgr~C} filament and
\objectname[NTF]{NTF~359.32$-$0.16} demonstrate that the GC magnetic
field cannot be dominated by a largely dipolar, poloidal component.

The new 90~cm survey coupled with the 6~cm observations described here indicate that the population of galactic center NTFs is significantly larger than previously thought.  These new NTFs can not be interpreted as components of a globally ordered poloidal magnetic field and suggest a more complicated field structure.    
Further, high sensitivity observations, especially with new instrumentation such as the expanded VLA and LOFAR (see Kassim et al 2003; White, Kassim \& Erickson 2003) will be required to discover
and characterize the entire population of Galactic center NTFs. 
         
\acknowledgments
We thank Harvey Liszt for sharing his 18~cm \objectname[]{Sgr~C} data and Steve Shore for many stimulating discussions concerning galactic center physics.  We also thank the referee M. Sakano for a thoughtful reading of the manuscript which significantly improved the paper.  Basic research in radio astronomy at the NRL is supported by the Office of Naval Research.

\clearpage

\begin{thebibliography}{}

\bibitem[\protect\citeauthoryear{Bicknell \& Li}{2001}]{bl01} Bicknell, G.\ \& Li, J. 2001, PASA, 18, 431

\bibitem[\protect\citeauthoryear{Bland-Hawthorn \&
	Cohen}{2003}]{b-hc03} Bland-Hawthorn, J.\ \& Cohen, M.  2003, \apj, 582, 246

\bibitem[\protect\citeauthoryear{Chakrabarti, Rosner, \&
	Vainshtein}{Chakrabarti et al.}{1994}]{crv94} Chakrabarti,
	S.~K., Rosner, R., \& Vainshtein, S.~I.  1994, Nature, 368, 434

\bibitem[\protect\citeauthoryear{Chandran}{2001}]{c01} Chandran,
	B.~D.~G.  2001, \apj, 562, 737

\bibitem[\protect\citeauthoryear{Chandran, Cowley, \&
	Morris}{2000}]{ccm00} Chandran, B.~D.~G., Cowley, S.~C., \&
	Morris, M.  2000, \apj, 528, 723

\bibitem[\protect\citeauthoryear{Chevalier}{1992}]{c92} Chevalier,
	R.~A.  1992, \apj, 397, 39


\bibitem[\protect\citeauthoryear{Chuss et al.}{2003}]{cnhdvdd03} Chuss, D.~T., Davidson, 
J.~A., Dotson, J.~L., Dowell, C.~D., Hildebrand, R.~H., Novak, G., \& 
Vaillancourt, J.~E.\ 2003, \apj, 599, 1116 

\bibitem[\protect\citeauthoryear{Crutcher}{1999}]{c99} Crutcher, R.~M.
	1999, \apj, 520, 706

\bibitem[\protect\citeauthoryear{Dahlburg et al.}{2002}]{dels02} Dahlburg, R.~B., Einaudi, G., LaRosa, T.~N., \& Shore, S.~N.,
	2002, \apj, 568, 220

\bibitem[\protect\citeauthoryear{Daly \& Loeb}{1990}]{dl90} Daly,
	R.~A.\ \& Loeb, A.  1990, \apj, 364, 451

\bibitem[\protect\citeauthoryear{Gregori et al.}{2000}]{gmrj00}
	Gregori, G., Miniati, M., Ryu, D., \& Jones, T.~W.  2000, \apj, 543, 775

\bibitem[\protect\citeauthoryear{Gray et al.}{1995}]{gnec95}Gray, A.~D., Nicholls, J., Ekers, R.D. \& Cram, L.~E., 1995, \apj, 448, 164

\bibitem[\protect\citeauthoryear{Heyvaerts, Norman, \&
	Pudritz}{1998}]{hnp98} Heyvaerts, J., Norman, C., \& Pudritz, R.~E. 1998, \apj, 330, 718

\bibitem[\protect\citeauthoryear{Kassim et al}{2003}]{kas03}Kassim, N.~E., Lazio, T.~J.~W., Nord, M., Hyman, S.~D., Brogan, C.~L., LaRosa, T.~N. \& Duric, N., Astron. Nachr. in press 2004

\bibitem[\protect\citeauthoryear{Koyama et al.}{1996}]{kmstty96} Koyama, K., Maeda, Y., Sonobe, T. Takeshima, T. Tanaka, Y., \& Yamauchi, S. 1996, \pasj, 48, 249\

\bibitem[\protect\citeauthoryear{Kronberg}{2002}]{k02} Kronberg, P.~P. 2002, Phys.\ Today, 55, 12, 40

\bibitem[\protect\citeauthoryear{Lang, Morris, \& Echevarria}{Lang et
	al.}{1999}]{lme99} Lang, C.~C., Morris, M.,
	\& Echevarria, L.  1999, \apj, 526, 727 

\bibitem[\protect\citeauthoryear{Lang et al.}{1999}]{laklg99} Lang, C.~C., Anantharamaiah,
	K.~R., Kassim, N.~E., Lazio, T.~J.~W., \& Goss, W.~M. 1999,
	\apj, 521, L41
	
\bibitem[\protect\citeauthoryear{LaRosa et al.}{2003}]{lnlk03} LaRosa,
	T.~N., Nord, M.~E., Lazio, T.~J.~W., S.~N. Shore \& Kassim, N.~E.  2004,
	Astron.\ Nachr., 324, in press
	
\bibitem[\protect\citeauthoryear{LaRosa et al.}{2000}]{lklh00} LaRosa, T.~N., Kassim, N.~E., Lazio, T.~J.~W., \&
	Hyman, S.~D. 2000, \aj, 119, 207

\bibitem[\protect\citeauthoryear{Liszt \& Spiker}{1995}]{ls95} Liszt, H.~S.\ \& Spiker, R. 1995, \apjs, 98, 259

\bibitem[\protect\citeauthoryear{Morris}{1998}]{m98} Morris, M. 1998,
	in The Central Regions of the Galaxy and Galaxies, ed.\
	Y.~Sofue (Dordrecht:Kluwer) p.~331

\bibitem[\protect\citeauthoryear{Morris}{1996}]{m96} Morris, M. 1996, in IAU Symposium 169,
	Unsolved Problems of the Milky Way, eds.\ L.~Blitz \& P.~Teuben 
	(Dordrecht: Kluwer) p.~247
	
\bibitem[\protect\citeauthoryear{Morris}{1994}]{m94} Morris, M. 1994,
	in The Nuclei of Normal Galaxies, eds.\ R.~Genzel \& D.~Harris
	(Dordrecht: Kluwer) p.~185

\bibitem[\protect\citeauthoryear{Morris \& Serabyn}{1996}]{ms96} Morris, M.\ \& Serabyn, E. 1996, \araa, 34, 645
	
\bibitem[\protect\citeauthoryear{Nord et al.}{2003a}]{nbhlklad03}
Nord, M.~E., Brogan, C.~L., Hyman, S.~D., Lazio, T.~J.~W., 
Kassim, N.~E., LaRosa, T.~N., Anantharamaiah, K., \& Duric,N. 2004, Astron.\ Nachr., 324, in press

\bibitem[\protect\citeauthoryear{Nord et al.}{2003b}]{nlkhld03} Nord, M.~E., Lazio, T.~J.~W., Kassim, N.~E., Hyman, S.~D., 
	LaRosa, T.~N., \& Duric, N.  2004, \aj, in preparation
	
\bibitem[\protect\citeauthoryear{Novak et al.}{2003}]{novaketal03} Novak, G., Chuss, D.~T., Renbarger, T., Griffin, G.~S., 
	Newcomb, M.~G., Peterson, J.~B., Loewenstein, R.~F., Pernic, D., \& Dotson, J.~L., 2003, \apj, 583, L83
	
\bibitem[\protect\citeauthoryear{Peirce-Price et
	al.}{2000}]{pierce-priceetal00} Pierce-Price, D., et al, 2000, \apj, 545, L121

\bibitem[\protect\citeauthoryear{Reich}{2003}]{r03} Reich, W.  2003, \aap, 401, 1023


\bibitem[\protect\citeauthoryear{Reid}{1993}]{r93} Reid, M.J. 1993, ARA\&A, 31, 345

\bibitem[\protect\citeauthoryear{Serabyn \& Morris}{1994}]{sm94} Serabyn, E. \& Morris, M.
	1994, \apj, 424, L91

\bibitem[\protect\citeauthoryear{Shore \& LaRosa}{1999}]{sl99} Shore, S.~N.\ \& LaRosa,
	T.~N. 1999, \apj, 521, 587

\bibitem[\protect\citeauthoryear{Sofue}{1985}]{s85} Sofue, Y.  1985, PASJ, 37, 697

\bibitem[\protect\citeauthoryear{Sofue \& Handa}{1984}]{sh84} Sofue, Y. \& Handa, T., 1984, Nature, 310, 568

\bibitem[\protect\citeauthoryear{Sofue \& Fujimoto}{1987}]{sf87}
	Sofue, Y.\ \& Fujimoto, M., 1987, Publ.\ Astron.\ Soc.\ Japan, 39, 843

\bibitem[\protect\citeauthoryear{Tsuboi et al.}{1986}]{tihtksk86} Tsuboi, M., Inoue, M., Handa, T., Tabara, H., Kato, T., Sofue,
	Y., \& Kaifu, N. 1986, \aj, 92, 818

\bibitem[\protect\citeauthoryear{Uchida, Sofue, \& Shibata}{Uchida et
	al.}{1985}]{uss85} Uchida, Y., Sofue, Y. \& Shibata, K. 1985, Nature, 317, 699

\bibitem[\protect\citeauthoryear{White, Kassim, \& Erickson}{2003}]{w03} White, S., Kassim, N.~E. \& Erickson, W.~C. 2003, SPIE, 4853, 111

\bibitem[\protect\citeauthoryear{Widrow}{2002}]{w02} Widrow, L.~M.
	2002, Rev.\ Mod.\ Phys., 74, 775

\bibitem[\protect\citeauthoryear{Yusef-Zadeh}{2003}]{y-z03} Yusef-Zadeh, F. 
	2003, \apj, 598, 325

\bibitem[\protect\citeauthoryear{Yusef-Zadeh, Morris, \& Chance}{Yusef-Zadeh et al.}{1984}]{y-zmc84} Yusef-Zadeh, F., Morris,
	M., \& Chance, D. 1984, Nature, 310, 55

\bibitem[\protect\citeauthoryear{Yusef-Zadeh \& Bally}{1989}]{y-zb89} Yusef-Zadeh, F.\ \& Balley J. 1989, in IAU 
	Symposium 136, The Center of the Galaxy, ed.\ M.~Morris, (Dordrecht:
	Kluwer) p.~243
	 
\bibitem[\protect\citeauthoryear{Yusef-Zadeh \& Morris}{1987a}]{y-zm87a} Yusef-Zadeh, F.\ \&
	Morris, M.  1987a, \apj, 322, 721
	 
\bibitem[\protect\citeauthoryear{Yusef-Zadeh \& Morris}{1987b}]{y-zm87b} Yusef-Zadeh, F.\ \&
	Morris, M.  1987b, \aj, 94, 1128

\bibitem[\protect\citeauthoryear{Yusef-Zadeh, Wardle, \& Parastaran}{Yusef-Zadeh et al.}{1997}]{y-zwp97} Yusef-Zadeh, F., Wardle,
	M., \& Parastaran, P. 1997, \apj, 475, L119
\end{thebibliography}
\end{document}